\documentclass[aip,apl,reprint,ttisuperscriptaddress,twocolumn]{revtex4}
\usepackage[dvips]{graphicx}
\usepackage{subfigure}
\usepackage{bm}
\usepackage{color}
\usepackage{amsmath,amssymb}
\begin{document}
\title{
Cross effect of magnetic field and charge current on antiferromagnetic dynamics
}
\author{Yuta Yamane$^1$, Olena Gomonay$^{2,3}$, Hristo Velkov$^2$, and Jairo Sinova$^{2,4}$}
\affiliation{$^1$Center for Emergent Matter Science (CEMS), RIKEN, Wako, Saitama 351-0198, Japan}
\affiliation{$^2$Institut f\"{u}r Physik, Johannes Gutenberg Universit\"{a}t Mainz,D-55099 Mainz, Germany}
\affiliation{$^3$National Technical University of Ukraine “KPI”, 03056 Kyiv, Ukraine}
\affiliation{$^4$Institute of Physics ASCR, v.v.i., Cukrovarnicka 10, 162 53 Praha 6, Czech Republic}
\date{\today}
\begin{abstract}
We theoretically examine a cross effect of magnetic field and charge current on antiferromagnetic domain wall dynamics.
Since antiferromagnetic materials are largely insensitive to external magnetic fields in general, charge current has been shown recently as an alternative and efficient way to manipulate antiferromagnets.
We find a new role of the magnetic field in the antiferromagnetic dynamics that appears when it is combined with charge current, demonstrating a domain wall motion in the presence of both field and current. 
We show that a spatially-varying magnetic field can shift the current-driven domain-wall velocity, depending on the domain-wall structure and the direction of the field-gradient.
Our result suggests a novel concept of field-control of current-driven antiferromagnetic dynamics.
\end{abstract}
\maketitle
%=========================================================================
%           Introduction
%=========================================================================
\section{Introduction}
In recent years, antiferromagnets (AFMs) are generating more attention due to their potential to play pivotal roles in spintronics applications\cite{review1,review2,review3}.
AFMs are robust against external magnetic fields, produce no or negligibly small stray fields, and exhibit faster magnetic dynamics compared to ferromagnets.
The insensitivity of AFMs to magnetic fields, however, may also indicate that an external magnetic field does not provide an efficient method to manipulate AFMs, a fact that has hindered active applications of AFMs in today's technology.
In the emergent field of antiferromagnetic spintronics, charge current is proving to be capable of offering promising ways to access the AFM dynamics, via the spin-transfer effect\cite{stt1,stt2,stt3,stt4,stt5,stt6,dw1,dw2,dw3,dw4,dw5,yamane,barker,ezawa,hristo} and the N\'{e}el spin-orbit torque\cite{nsot1,nsot2,nsot3};
e.g., current-driven motion of AFM textures such as domain walls\cite{dw1,dw2,dw3,dw4,dw5,yamane,nsot3} and skyrmons\cite{barker,ezawa,hristo} have been proposed.

However, it may be too hasty to conclude that the magnetic field will not find its place in future spintronics applications.
The equation of motion for a two-sublattice AFM is a second order differential equation of time\cite{andreev,bary}, where an external magnetic field ${\bm H}$ and a charge current density ${\bm v}$ (in the unit of velocity\cite{note3}) enter as the factors $\gamma {\bm H}$, with $\gamma$ the gyromagnetic ratio, and $ ( {\bm v} \cdot \nabla ) $\cite{dw2,dw3,dw4,dw5,yamane}, respectively, each being in the unit of $t^{-1}$.
The AFMs therefore can allow for cross terms of magnetic field and charge current to appear directly in their equation of motion\cite{yamane}, unlike the ferromagnetic counterpart.
%To see this, let us take a look at one of the major differences between the equation of motion for two-sublattice AFM order vector and the ferromagnetic Landau-Lifshitz-Gilbert (LLG) equation, i.e., the former is a second-order differential equation of time\cite{andreev,bary} while the latter is first-order.
%In the LLG equation, an external magnetic field ${\bm H}$ and a charge current density ${\bm v}$ (in the unit of velocity) appear as the factors $\gamma {\bm H}$ and $ ( {\bm v} \cdot \nabla ) $, respectively, with $\gamma$ being the gyromagnetic ratio;
%since each factor is already in the unit of $t^{-1}$, there appears no cross term of field and current that would be higher order.
%For the AFMs, on the other hand, the order argument allows for the equation to contain such field-current cross terms.
The magnetic field may thus be able to play some roles in the AFM dynamics when it is combined with charge current.
Very recently, the equation of motion that contains such cross terms has been indeed derived for certain classes of two-sublattice AFMs\cite{yamane}.
It remains to be examined, however, how the cross terms manifest themselves and make impacts in concrete and practical systems.

In this work, we theoretically demonstrate a cross effect of external magnetic field and charge current on AFM domain wall (DW) dynamics in a thin nanowire.
To specify the effective spin-transfer effect, we restrict ourselves to a class of AFMs where the inter-sublattice electron transport is strongly suppressed.
We derive an equation of motion of the DW based on a collective-coordinate model, in the presence of both charge current and magnetic field.
It is shown that a spatially-varying magnetic field applied in the out-of-plane, which cannot drive the DW into motion by itself, either increases or decreases the current-driven DW velocity depending on the DW structure and the direction of field-gradient. 
Our results suggest the possibility of a novel way to manipulate AFMs, namely, field-control of the current-driven AFM dynamics.

%=========================================================================
%           Model
%=========================================================================
\section{Model}
We consider a thin nanowire of metallic AFM, which is composed of two sublattices (1 and 2) with equal saturation magnetization $M_{\rm S}$.
We employ the one-dimensional model along the $z$-axis, where we assume the uniformity of the magnetizations in the lateral directions, i.e., the $x$-$y$ plane.
(See Fig.~1 for our coordinate system.)
In order to treat the magnetizations classically, the coarse graining for the magnetic channel is performed\cite{neel,lifshitz}.
The classical vector $ {\bm m}_1 ( z , t ) $ $ ( | {\bm m}_1 ( z , t ) | = 1 ) $ is a continuous function in space that represents the local magnetization direction in the sublattice 1, with a similar definition for $ {\bm m}_2 ( z , t ) $;
here the lattice structure is smeared out and the magnetizations of both sublattices are defined at every point in space.
This classical treatment is allowed when the spatial variation of each magnetization is sufficiently slow compared to the atomistic length scale.

As more experimentally relevant quantities, we here introduce the ferromagnetic canting vector $ {\bm m} ( z , t ) = [ {\bm m}_1 ( z , t ) + {\bm m}_2 ( z , t ) ] / 2 $ and the N\'{e}el order vector $ {\bm n} ( {\bm r} , t ) = [ {\bm m}_1 ( z , t ) - {\bm m}_2 ( z , t ) ] / 2 $.
The conditions of $ {\bm m}^2 + {\bm n}^2 = 1 $ and $ {\bm m} \cdot {\bm n} = 0 $ are a direct consequence of the definitions of ${\bm m}$ and ${\bm n}$ given above.

We employ the following magnetic energy density $w ( z ) $ to describe the AFM;
\begin{equation}
w ( z )  =  w_{\rm exc} ( z ) + w_{\rm ani} ( z ) + w_{\rm zmn} ( z ) .
\end{equation}
The first term, $ w_{\rm exc} = \mu_0 M_{\rm S} H_E ( {\bm m}^2 - {\bm n}^2 ) + A ( \partial_z {\bm n} )^2 + A' ( \partial_z {\bm m} )^2 $, describes the exchange interactions between the local magnetizations, with $H_E$ representing the homogeneous exchange field, and $A$ and $A'$ being the inhomogeneous exchange constants.
The second term, $ w_{\rm ani} = K_y ( m_y^2 + n_y^2 ) - K_z ( m_z^2 + n_z^2 ) $, is the magnetic anisotropy energy, where the easy and hard-axes along the $z$ and $y$ directions, respectively, are assumed, with the anisotropy constants $ K_y ( > 0 ) $ and $ K_z ( > 0 ) $.
The third term, $w_{\rm zmn} = 2 \mu_0 M_{\rm S} {\bm H} \cdot {\bm m} $, is the Zeeman energy.
In this study we assume that the AFM exchange coupling between ${\bm m}_1$ and ${\bm m}_2$ is the leading energy scale, as usually is the case, being strong enough to ensure $| {\bm m} ( z, t ) | \ll 1$.

The equations of motion for ${\bm n}$ and ${\bm m}$ can be obtained by assuming two coupled Landau-Lifshitz-Gilbert equations for the sublattice magnetizations, where ${\bm H}^{\rm eff}_1 = - ( \mu_0 M_{\rm S} )^{-1} \delta w / \delta {\bm m}_1$ and ${\bm H}^{\rm eff}_2 = - ( \mu_0 M_{\rm S} )^{-1} \delta w / \delta {\bm m}_2$ act as the effective magnetic fields on ${\bm m}_1$ and ${\bm m}_2$, respectively, and then reading these equations in terms of ${\bm n}$ and ${\bm m}$.

In the presence of charge current, the effective interaction between the conduction electrons and the local magnetizations depends on the detail of the electron transport through the two sublattices\cite{dw5,yamane,barker}.
For the present purpose to explore the possibility of cross effects by charge current and magnetic field, we here focus on a simple limiting case where the inter-sublattice electron transport is virtually suppressed\cite{barker,yamane};
this is the case when the local exchange coupling between the conduction electron spin and the magnetization is strong compared to the electron's kinetic energy corresponding to the inter-sublattice hopping.
In this case, the equation of motion for ${\bm n}$ including the spin-transfer effect\cite{note} can be derived in an analytical and quantitative fashion as\cite{yamane}
\begin{eqnarray}
{\bm n} &\times& \left[ \left( {\cal D}^2_t - c^2 \partial_z^2 \right) {\bm n} + \gamma^2 \left( {\bm n} \cdot {\bm H} \right) {\bm H} + \gamma {\bm n} \times {\cal D}_t {\bm H} \right. \nonumber \\ && \left.
                                   - 2 \gamma \left( {\bm n} \cdot {\bm  H} \right) {\bm n} \times {\cal D}_t {\bm n}  +  \frac{ \gamma \left( K_z n_z \hat{{\bm z}} - K_y n_y \hat{{\bm y}} \right) }{ \mu_0 M_{\rm S} }   \right. \nonumber \\ && \left. 
                                   + 2 \gamma H_E \left( \alpha \partial_t - \beta u \partial_z \right) {\bm n} 
                                   \right] = 0  ,
\label{n} \end{eqnarray}
while $ {\bm m} $ is determined as a function of $ {\bm n} $;
\begin{equation}
{\bm m}  =  \frac{ 1 }{ 2 \gamma H_E } \left( {\cal D}_t {\bm n} + {\bm n} \times \gamma {\bm H}  \right) \times {\bm n} , 
\label{m} \end{equation}
where $\alpha$ and $\beta$ are dimensionless parameters describing the dissipative process, and $c^2 =  2 \gamma^2 H_E / \mu_0 M_{\rm S} $.
The spin-transfer effects brought on by the charge current are reflected in the Lagrange derivative ${\cal D}_t$ that is defined by
\begin{equation}
{\cal D}_t  =  \partial_t - u \partial_z  ,
\label{dt} \end{equation}
where $ u = ( g \mu_B P_{\rm sub} / 2 e M_{\rm S} ) j_{\rm c} $, with $g$ the g factor, $\mu_B$ the Bohr magneton, $e$ the elementary charge, $j_c$ the charge current density, and $P_{\rm sub}$ the spin polarization defined in {\it each sublattice}\cite{note2}.
In deriving Eqs.~(\ref{n}) and (\ref{m}) we have used the conditions $ \mu_0 M_{\rm S} H_E \gg K_y, K_z $, $ | {\bm m} | \ll 1 $, and $ {\bm n}^2 \simeq 1 $. 
Eqs.~(\ref{n}) and (\ref{m}) respect the Galilean invariance with respect to the charge current when $ \alpha = \beta = 0 $.

The third and forth terms in Eq.~(\ref{n}) contain both ${\cal D}_t$ and ${\bm H}$, i.e., they are cross terms of charge current and magnetic field.
Although the possibility of field-current cross effects due to these terms was already pointed out\cite{yamane}, it remains to be confirmed in concrete and practical systems.
In the following, we demonstrate domain wall (DW) motion in the presence of uniform dc charge current and spatially-varying dc magnetic field, where the third term in Eq.~(\ref{n}) plays a role.

\begin{figure}
  \centering
  \includegraphics[width=8cm,bb=0 0 790 540]{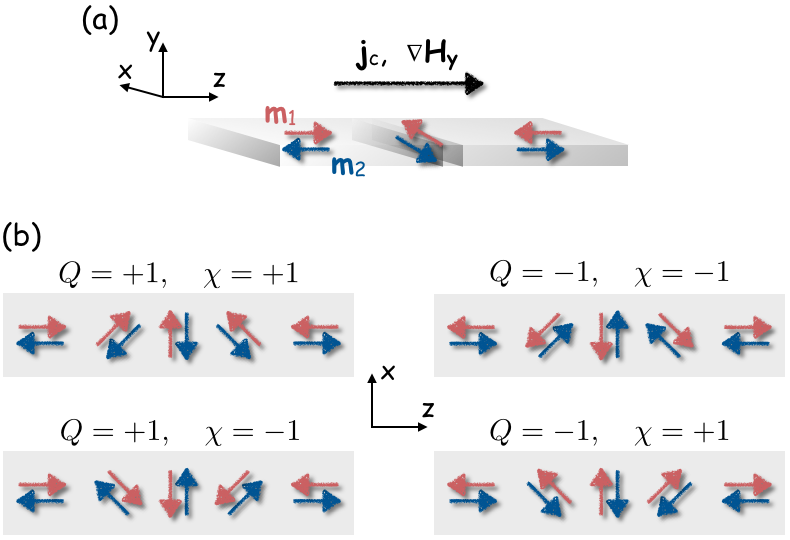}
  \caption{ {\bf a}. Schematic of our system.
                 A one-dimensional domain wall is formed in the antiferromagnetic nanowire that is composed of the two sublattice-magnetizations ${\bm m}_1$ and ${\bm m}_2$.
                 In order to drive the DW into the motion, the charge current ${\bm j}_c$ and the magnetic field ${\bm H}$ are applied along the $z$ and $y$ axes, respectively, with the field-gradient realized in the $z$ direction.
                 {\bf b}. Schematics of domain wall configurations with different sets of ($Q$, $\chi$).
                 (see the main text for the definitions of the quantities.)
                 Because of the equivalence of the two sublattices, ($+1,+1$) and ($-1,-1$) [($+1,-1$) and ($-1,+1$)] are indistinguishable, each being characterized by $Q \chi = +1$ [$-1$]. 
               }
  \label{fig01}
\end{figure}

%=========================================================================
%           Domain Wall Motion
%=========================================================================
\section{Domain wall motion}
In equilibrium with no magnetic field, it is seen from Eq.~(\ref{m}) that $ {\bm m} = 0 $.
A one-dimensional DW solution for ${\bm n} ( z ) = ( \sin\theta ( z ) \cos\phi ( z ) , \sin\theta ( z ) \sin\phi ( z ) , \cos\theta ( z ) ) $ is obtained by locally-minimizing the magnetic energy $ \sigma \equiv \int_{-\infty}^\infty w \ dz $ with the boundary condition $ \theta ( \pm \infty) = ( 0 , \pi ) $ or $ ( \pi , 0 ) $;
\begin{eqnarray}
\theta \left( z \right)  &=&  2 \tan^{-1} \left[ \exp \left( Q \frac{ z - q }{ \Delta } \right) \right]  , \label{theta} \\
\chi  &\equiv&  \cos\phi  =  \pm 1 ,  \label{phi}
\end{eqnarray}
where $q$ represents the position of the DW center, $Q$ is the topological charge of the DW;
\begin{equation}
Q  =  \frac{ 1 }{ \pi } \int_{ - \infty }^\infty  \partial_z \theta dz  =  \pm 1 ,
\end{equation}
and $\Delta$ is the DW-width parameter defined by
\begin{equation}
\Delta = \sqrt{ \frac{ A }{ K } }  .
\label{delta} \end{equation}
Schematics of the DW configurations with different sets of $ ( Q , \chi ) $ are shown in Fig.~1~(b).
Because of the equivalence of the two sublattices, ($ + 1 , + 1 $) and ($ - 1 , - 1 $) are indistinguishable,  and so are ($ + 1 , - 1 $) and ($ - 1 , + 1 $);
the DW can be characterized by $ Q \chi = \pm1 $.
Eq.~(\ref{theta}) is plotted in the inset of Fig.~2.

\begin{figure}[t]
  \centering
  \includegraphics[width=8cm,bb=0 0 672 487]{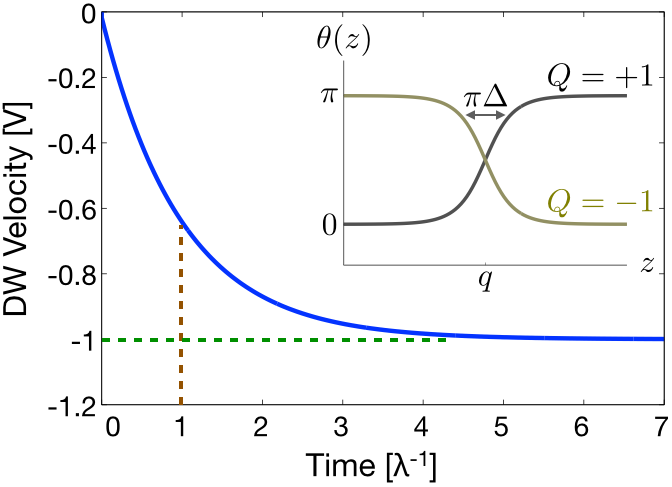}
  \caption{ Blue curve;
                 The domain-wall velocity $ \partial_t q $ [in the unit of $V$, Eq.~(\ref{v})] as a function of time [in the unit of $\lambda^{-1}$, Eq.~(\ref{lambda})], based on Eq.~(\ref{dq}) with initial condition $ \left. \partial_t q \right|_{ t = 0 }  = 0 $.
                 The green dotted line indicates the terminal velocity, where $V$ is assumed to be negative.
                 The characteristic time for the velocity relaxation is indicated by the brown dotted line.
                 Inset; Eq.~(\ref{theta}) is plotted as a function of $z$ with $ Q = \pm 1 $.
                 The domain-wall width parameter $\Delta$ is defined in Eq.~(\ref{delta}).
                 In the present model, this DW structure is assumed to be sustained in the presence of current and field as well as in equilibrium, save for the time dependence of $ q ( t ) $.
                }
  \label{fig02}
\end{figure}

A charge current and magnetic field can drive the DW into motion according to Eq.~(\ref{n}).
Here we assume that the driving forces due to the current and field are weak enough that the DW sustains its structure given by Eqs.~(\ref{theta}) and (\ref{phi}), except that the DW center $ q ( t ) $ becomes time-dependent;
the DW dynamics is described by the time evolution of the collective coordinate $ q ( t ) $.
In the presence of uniform dc charge current flowing along the nanowire (the $z$ axis) and spatially-varying dc magnetic field applied in the our-of-plane (the $y$ axis), Eq.~(\ref{n}) with the above ansatz is reduced to
\begin{equation}
\partial_t^2 q  + \lambda \partial_t q  =  - 2 u \beta \gamma H_E  +  Q \chi \frac{ \pi \gamma \Delta }{ 2 } u \partial_z H  ,
\label{eom} \end{equation}
where $\partial_z H$ is assumed to be constant, and
\begin{equation}
\lambda = 2 \alpha \gamma H_E .
\label{lambda} \end{equation}
In the right-hand side of Eq.~(\ref{eom}), we find the two driving forces on $ q ( t ) $;
the first term is solely by the charge current, having its origin in the last term in Eq.~(\ref{n}), whereas the second term is a cross term of the current and field, whose origin is the third term in Eq.~(\ref{n}).
Notice that the sign of the cross term depends on $ Q \chi = \pm 1 $.

\begin{figure}[t]
  \centering
  \includegraphics[width=8cm,bb=0 0 910 511]{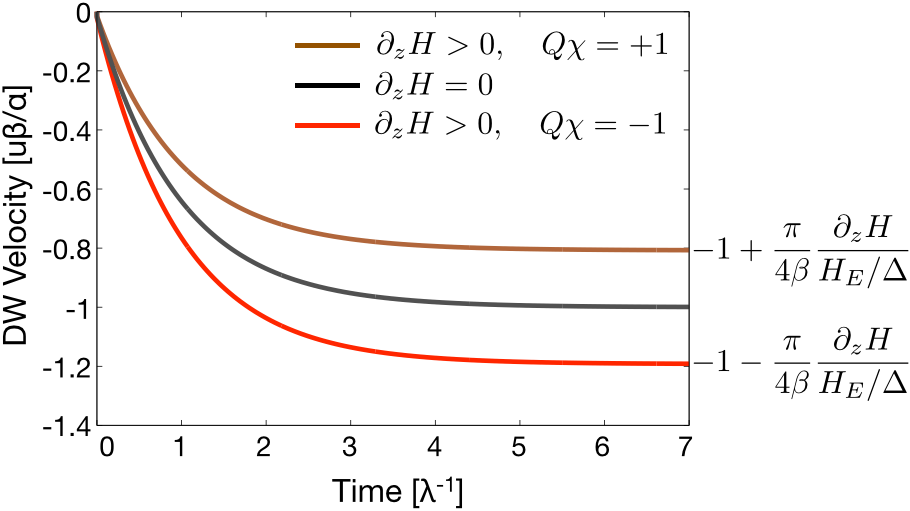}
  \caption{ The domain-wall velocity $ \partial_t q $ (in the unit of $ u \beta / \alpha$) is plotted as a function of time (in the unit of $\lambda^{-1}$), with different field-gradients.
                 With no field-gradient (black curve), the velocity reaches $ - 1 $, as is known.
                 The presence of field-gradient $\partial_z H $ (red and brown curves) leads to either increase or decrease in the DW velocity, depending on $ Q \chi $.
                 (Here $\partial_z H $ is assumed to be positive.)
                 See the main text for the definitions of the other quantities.
                }
  \label{fig03}
\end{figure}

General solutions,  $ \partial_t q ( t ) $ and $ q ( t ) $, of Eq.~(\ref{eom}) are obtained as (see Fig.~2)
\begin{equation}
\partial_t q ( t )  =  V \left( 1 - e^{ - \lambda t } \right) + \left. \partial_t q \right|_{ t = 0 } e^{ - \lambda t }  ,
\label{dq} \end{equation}
\begin{equation}
q ( t )  =  V t  + \frac{ V - \left. \partial_t q \right|_{ t = 0 } }{ \lambda } e^{ - \lambda t } + q( 0 ) - \frac{ V - \left. \partial_t q \right|_{ t = 0 } }{ \lambda }  ,
\end{equation}
where
\begin{equation}
V  \equiv  \left. \partial_t q \right|_{ t \rightarrow \infty }  =  - \frac{ u }{ \alpha } \left( \beta - \frac{ \pi Q \chi }{ 4 } \frac{  \partial_z H }{ H_E / \Delta} \right)  .
\label{v} \end{equation}
In the absence of field gradient, the terminal velocity $V$ is proportional to the ratio $ - \beta / \alpha $ of the dissipative parameters, being consistent with the existing study\cite{dw2,dw3,dw4,yamane}.
The coexistence of the magnetic field and charge current leads to either increase or decrease in $V$, depending on the direction of the field-gradient and $Q \chi = \pm 1$ (see Fig.~3).

%For antiferromagnetic materials, there is so far little (or almost no) experimental data available on the values of $\alpha$ and $\beta$, which are crucial for determination of the DW dynamics.
%We here try to give an estimation of the field-current cross effect by employing $ \alpha = 0.01 $ and $\beta = 0.02$, which are the typical values for ferromagnets, and $ \gamma = 1.76 \times 10^{11} $ Hz/T, $ \Delta = 30 $ nm and $ \lambda = 4.4 \times 10^{10} $ s$^{-1}$, which are in the reasonable range for typical AFMs.
%With field-gradient of $ \gamma \partial_z H = $
%=========================================================================
%           Conclusions
%=========================================================================
\section{Discussions and Conclusions}
Our attempt to estimate the terminal velocity $V$ faces the difficulty that for most AFMs there is only little or almost no experimental data available on the values of the crucial material parameters such as $\alpha$ and $\beta$.
Employing $ \mu_0 H_E = 10 $ T and $\Delta = 100$ nm, which are in the reasonable range for typical AFMs\cite{gurevich}, the ratio of the absolute values of the first and second terms in Eq.~(\ref{v}) is given by
\begin{equation}
\frac{ 1 }{ \beta } \left| \frac{ \pi Q \chi }{ 4 } \frac{  \partial_z ( \mu_0 H ) }{ \mu_0 H_E / \Delta} \right|  \simeq 0.79 \times 10^{-8} \frac{ | \partial_z ( \mu_0 H ) | }{ \beta }
\label{est} \end{equation}
Assuming $ | \partial_z ( \mu_0 H ) | = 100 $ T/m, and borrowing the typical value $\beta = 0.01 $ of the nonadiabatic parameter for the ferromagnets, Eq.~(\ref{est}) is evaluated as $\sim 10^{-4}$;
in typical AFMs the field-current cross term in Eq.~(\ref{v}) is expected to be small compared to the other term.
To identify the proposed field-current cross effect on DW dynamics, therefore, material should be chosen carefully.
The presence of field-gradient would lead to a visible shift in the current-driven DW velocity when smaller $\beta$ and $H_E$, and larger $\Delta$ (that is, smaller anisotropy constant $K_z$) are realized.

Effects of an inhomogeneous magnetic field was also investigated in Ref.~\cite{tveten}.
It was proposed that a spatially-varying external magnetic field, applied along the nanowire unlike our case, can drive a DW motion even in the absence of charge current.
Their argument is based on the observation that the spatial variation of the N\'{e}el order vector around the DW is accompanied by finite ferromagnetic canting moment, which directly couples to the external magnetic field.
This effect is emphasized when the DW width becomes as small as a few lattice spacings;
we have neglected this effect because our model assumes the slow spatial variation of the magnetizations compared to the atomistic length scale.
At any rate, the authors of Ref.~\cite{tveten} and we look at different effects, which may be compatible with each other.

While we have for simplicity restricted ourselves to the limiting case where the inter-sublattice electron transport is suppressed, the equations of motion (\ref{n}) and (\ref{eom}) would not simply apply to other classes of AFMs\cite{yamane};
the role played by a magnetic field in the DW dynamics should be reformulated and reexamined there.
Lastly, while the other field-current cross term, the forth term in Eq.~(\ref{n}), can be neglected in the one-dimensional DW dynamics, it would not be the case in more complex structures. 
Detailed investigations into these quenstions would be possible directions for further research.

In conclusion, we have theoretically studied antiferromagnetic domain wall dynamics in the presence of charge current and magnetic field.
It has been shown that a spatially-varying magnetic field can either increase or decrease the current-driven domain-wall velocity depending on the domain wall structure and the direction of the field-gradient.
This is the first demonstration of a field-current cross effect on antiferromagnetic dynamics.
We believe that our result has made an important step towards the field-control of the current-driven antiferromagnetic dynamics.

%=========================================================================
%           Acknowledgments
%=========================================================================
This research was supported by Research Fellowship for Young Scientists from Japan Society for the Promotion of Science, the Alexander von Humboldt Foundation, the ERC Synergy Grant SC2 (No.~610115), the Transregional Collaborative Research Center (SFB/TRR) 173 SPIN+X, and Grant Agency of the Czech Republic grant No.~14-37427G.

%%%%%%%%%%%%%%%%%%%%%%%%%%%%%%%%%%%%%%%%%%%%%%%%%
%%%%%%%%%%%%%%%%%%%%%%%%%%%%%%%%%%%%%%%%%%%%%%%%%
%%%%%%%%%%%%%%%%%%%%%%%%%%%%%%%%%%%%%%%%%%%%%%%%%
%%%%%%%%%%%%%%%%%%%%%%%%%%%%%%%%%%%%%%%%%%%%%%%%%
%%%%%%%%%%%%%%%%%%%%%%%%%%%%%%%%%%%%%%%%%%%%%%%%%
%%%%%%%%%%%%%%%%%%%%%%%%%%%%%%%%%%%%%%%%%%%%%%%%%
%%%%%%%%%%%%%%%%%%%%%%%%%%%%%%%%%%%%%%%%%%%%%%%%%
%%%%%%%%%%%%%%%%%%%%%%%%%%%%%%%%%%%%%%%%%%%%%%%%%
%%%%%%%%%%%%%%%%%%%%%%%%%%%%%%%%%%%%%%%%%%%%%%%%%
%%%%%%%%%%%%%%%%%%%%%%%%%%%%%%%%%%%%%%%%%%%%%%%%%
%%%%%%%%%%%%%%%%%%%%%%%%%%%%%%%%%%%%%%%%%%%%%%%%%
%=========================================================================
%           References
%=========================================================================


\begin{thebibliography}{99}
\bibitem{note3}  The explicit expression for ${\bm v}$ may depend on the system.
                          For the class of AFMs considered in this work, $ | {\bm v} | = u $ is defined right below Eq.~(\ref{dt}).
\bibitem{note}  In the present study, we don't consider the N\'{e}el spin-orbit torque because it can be present only in materials that possess special symmetry\cite{nsot1,nsot2,nsot3}.
\bibitem{note2}  $P_{\rm sub}$ does {\it not} represent the net spin polarization of the bulk sample, which is zero in the present case.
                         Rather, $P_{\rm sub}$ describes the fact that the majority and minority bands are split in each sublattice by $ \sim 2 J $, where $J$ is the exchange coupling energy between the conduction electron spin and the local magnetization.
\end{thebibliography}
\end{document}